# A simulation and case study to evaluate the extrapolation performance of flexible Bayesian survival models when incorporating real-world data

Iain R. Timmins[1]*, Fatemeh Torabi[2-4], Christopher H. Jackson[4], Paul C. Lambert[5,6], Michael J. Sweeting[1,7]

[1]Statistical Innovation, Oncology R&D, AstraZeneca, Cambridge, UK

[2]Institute of Continuing Education, University of Cambridge, Cambridge, UK

[3]Dementias Platform UK, Swansea University, Swansea, Wales, UK

[4]MRC Biostatistics Unit, University of Cambridge, Cambridge, UK

[5]Cancer Registry of Norway, Norwegian Institute of Public Health, Oslo, Norway

[6]Department of Medical Epidemiology and Biostatistics, Karolinska Institutet, Sweden

[7]Department of Population Health Sciences, University of Leicester, UK

*email: iain.timmins@astrazeneca.com

# Abstract

*Background*

Assessment of long-term survival for health technology assessment often necessitates extrapolation beyond the duration of a clinical trial. Without robust methods and external data, extrapolations are unreliable. Flexible Bayesian survival models that incorporate longer-term data sources, including registry data and population mortality, have been proposed as an alternative to using standard parametric models with trial data alone.

*Methods*

The accuracy and uncertainty of extrapolations from the `survextrap` Bayesian survival model and R package were evaluated. In case studies and simulations, we assessed the accuracy of estimates with and without long-term data, under different assumptions about the long-term hazard rate and how it differs between datasets, and about treatment effects.

*Results*

The `survextrap` model gives accurate extrapolations of long-term survival when long-term data on the patients of interest are included. Even using moderately biased external data gives improvements over using the short-term trial data alone. Furthermore, the model gives accurate extrapolations of differences in survival between treatment groups, provided that a reasonably accurate assumption is made about how the treatment effect will change over time. If no long-term data are available, then the model can quantify structural uncertainty about potential future changes in hazard rates.

*Limitations*

The conclusions from the case study and simulation study are based on a specific trial and real-world data structure and data-generating mechanism. The quantification of long-term structural uncertainty is approximate in the absence of long-term data.

*Conclusions*

This analysis shows that Bayesian modelling can give accurate and reliable survival extrapolations by making the most of all available trial and real-world data. This work improves confidence in the use of a powerful tool for evidence-based healthcare decision-making.

# Introduction

Survival extrapolation is vital in health technology assessment (HTA) and cost-effectiveness analysis, where long-term outcomes are needed for informed decision-making. The limited time frame of clinical trials often fails to capture long-term treatment effects and survival patterns, which can result in substantial uncertainty and potential bias when predicting future outcomes. Without robust methods and external data to bridge this gap, decision-makers face difficulties in accurately assessing the cost-effectiveness of new interventions. Several authors have suggested using external data or real-world evidence (RWE)[1–5], such as registries, population mortality rates from life tables, and expert elicited opinion, to be incorporated into models to better anchor extrapolations[1–12].

Recently, a novel Bayesian evidence synthesis model for survival data has been presented in the R package 'survextrap'[6]. This tool facilitates survival extrapolation by allowing the user to incorporate multiple data sources, with the hazard modelled as a flexible spline function. The Bayesian approach can represent uncertainty about the extrapolated survival function, ensuring that extrapolations are only confident if there is long-term data to support them. While Jackson[6] presented a case study of the use of the 'survextrap' package, there was no formal assessment of the accuracy and robustness of the models. As this is a general-purpose and flexible tool for evidence synthesis, appropriate modelling involves many choices. In this paper, we aim to highlight and guide the most important choices that need to be explicitly considered by users of this tool, and to inform "default" settings for less important assumptions. Previously[13] we studied its use for short-term estimation during the trial follow-up, while in this paper we focus on assumptions that affect long-term extrapolation of both survival and treatment effects.

In particular, we focus on how to model external data, which are generally included to inform long-term survival for the trial control population, or background mortality rates for all patients. However, external data may be subject to unobserved confounding, information biases in real-world sources[14], and differences between trial and real-world populations. Moreover, the appropriate level of flexibility that the model needs to represent data of this form is unclear. We also assess how the long-term treatment effect is modelled, since there will generally be little to no data on the long-term effect of a novel treatment, therefore we require the use of further parametric assumptions and sensitivity analyses. Firstly, we present a case study (similar to that in Jackson[6] and Guyot[9]) that demonstrates, in the presence and absence of external data, the potential modelling choices available and their impact. Secondly, we perform a simulation study to demonstrate the relative accuracy on long-term survival metrics of different modelling choices using trial and external data generated from a plausible survival and treatment model. Through these aims we seek to better understand the value of including longer-term real-world evidence to anchor survival extrapolations using a flexible Bayesian approach, helping to guide practitioners on each aspect of model selection, application of sensitivity analyses (e.g., treatment effect waning), and transparent assessment of extrapolation uncertainty.

# Methods

## The `survextrap` R package

The '`survextrap`' R package fits a Bayesian evidence synthesis to multiple sources of survival data through the specification of a flexible parametric survival model[6]. The Bayesian model can flexibly accommodate an individual-level dataset for the clinical trial under consideration and one or more aggregate external datasets giving counts of survivors over arbitrary time periods. A single hazard function is assumed to describe all sources of data, whilst heterogeneity between datasets can be modelled through covariates. Furthermore, it can integrate general population mortality rates through an excess hazard (relative survival) framework[3], while cure models are also supported. For modelling treatment effects, a proportional hazards (PH) or a flexible non-proportional hazards (Non-PH) model can be used, as well as modelling treatment arms separately, and treatment effect waning scenarios can be investigated through sensitivity analysis.

**A flexible M-spline to model the hazard function**

The model uses an M-spline to describe the hazard function, which is defined as a weighted sum of basis functions that are cubic by default, resulting in a smooth, positive-valued function. The flexibility of the spline is determined by the number of basis functions and corresponding knot locations, and the choice of priors on the scale parameter $\eta$ (the overall magnitude of the hazard) and a smoothness parameter $\sigma$. Further details are given in the Supplementary Material.

**Modelling 'structural uncertainty' through knot placement**

A key feature of the M-spline hazard function (or excess hazard if using relative survival) is that it is assumed to be constant after the last knot location, which by default will be at or close to the end of the trial follow-up. Therefore, to allow flexible modelling of longer-term external data, we may place extra knots after the end of trial follow-up. There should be enough long-term knots to capture any variations over time observed in the external data, and any variations that are expected in the absence of observed data. The trial data will then influence the long-term extrapolation only through the assumption that the hazard (or excess hazard) changes smoothly through time, and as such, the M-spline can account for the complete "structural uncertainty" about how the hazard may change after the trial data. An illustrative example of `survextrap` models accounting for structural uncertainty is provided in Figure 1.

In conventional parametric extrapolation, the hazard function over all times is described by parameters learnt from short-term data alone, which neglects uncertainty about how the parameter values (or the parametric form itself, e.g. Weibull or Gamma) might change in the long term. Averaging over alternative parametric extrapolations has been suggested as a way to account for such structural uncertainty[15]. However, none of those extrapolations will necessarily represent the overall hazard curve well, and the appropriate choice of averaging weight is unclear.

M-splines provide a more elegant, data-adaptive approach. They are arbitrarily flexible, and each parameter describes only the hazard in the region of time surrounding a particular knot. Therefore if extra knots outside the trial follow-up are added, then the parameters relating to the short-term knots (estimated from short-term data) will have a negligible influence on the long-term trajectory. If there are no long-term data, the uncertainty about the long-term parameters will be appropriately high, and any "extrapolation" from short-term data will then be based only on the modest and plausible assumption that the hazard function is varying smoothly.

## Case Study

As a motivating example, we reanalysed survival extrapolations based on a trial with 5 years of follow-up comparing cetuximab and radiotherapy to radiotherapy alone in patients with head and neck cancer[16], which was previously examined by Jackson[6] and Guyot[9]. The aim was to assess the impact of modelling choices, in the presence and absence of external data, on extrapolated survival up to a lifetime horizon of 40 years after randomization.

Patient-level time-to-event trial data were obtained from digitised published 5-year Kaplan-Meier curves for overall survival (OS)[16,17]. In addition to the trial data, we used two external data sources previously curated by Guyot et al.[9]. The first used registry data from the US National Cancer Institute Surveillance, Epidemiology, and End Results (SEER) database, where patients were matched to the trial population by age, gender, cancer site, and date of diagnosis. Annual counts of survivorship were provided from years 5 to 25 after diagnosis and considered representative of the long-term survival rates of the radiotherapy control arm. A second source of data was general population mortality rates from US lifetables, which were matched to the summary level characteristics of the trial (80% male, median age 57 at diagnosis) since individual covariate information on age and sex were not available from the trial data.

### Bayesian survival models

Extrapolation beyond the duration of the trial requires careful assessment of how different modelling choices impact long-term survival predictions. We considered three modelling choices: (a) position of the long-term knots, (b) how the external data are modelled, and (c) how the treatment effect is modelled.

For the position of long-term knots, beyond the follow-up of the trial, we compare the impact of four alternative scenarios: 1) no additional knots beyond trial follow-up, 2) one additional knot at 25 years, 3) two additional knots at 10 and 25 years, and 4) three additional knots at 10, 15 and 25 years. These knots are chosen to span a period where a non-negligible proportion of patients remain alive and hence modelling choices will be important.

Regarding the choice of how external data are modelled, , we considered models with 1) trial data alone, 2) trial data plus general population mortality rates (represented via an excess hazard model), and 3) trial data, population mortality rates, and SEER registry data on the control arm.

Utilising all three data sources, we modelled the treatment effect using PH, Non-PH and separate-arm models, which make different assumptions about how the treatment and control groups are related in the short and long term. These models are used in conjunction with different assumptions about whether and how the treatment effect will wane over time. Without any treatment effect waning, under the PH model, the treatment effect is extrapolated by making a strong assumption that the (excess) hazard ratio is constant after the trial data. Details of the priors, treatment waning, and other modelling choices used in the Bayesian models are provided in the Supplementary Material.

# Simulation Study

To formally investigate the impact of modelling choices on the bias and precision of long-term survival estimates, we performed a simulation study.

## Data Generating Mechanism

We simulated survival times for a two-arm trial with 1:1 randomisation and sample size 400 (200 per arm), with follow-up of 5 years, generating 1,000 trial datasets. We additionally considered scenarios with 3- and 8-years follow-up to assess the impact of data cut-off. The ages of trial participants were simulated from a normal distribution with mean 60 years and standard deviation 9 years. We modelled our control arm to fit the overall survival of patients with head and neck cancer treated with radiotherapy from the Bonner et al. trial[16].

Survival times for the control arm were generated from an additive hazards model that included both disease specific and other-cause hazards. The disease-specific hazard was defined from a Weibull mixture distribution, and the other-cause hazard function was a Gompertz distribution with age as the time scale with Gompertz parameters based on English mortality rates[18]. Further details are given in the Supplementary Material.

To generate data for the active arm we considered three treatment effect scenarios: a constant disease-specific treatment effect giving rise to proportional excess hazards (scenario 1), an immediate treatment effect followed by treatment effect waning (scenario 2) and a delayed treatment effect followed by treatment effect waning (scenario 3). The functions are provided in Supplementary Table 2. Random uniform censoring was applied to the last two years of trial follow-up (years 3-5), to mimic staggered recruitment into the trial.

External data was generated to inform control arm hazard rates from years 6 to 25 (i.e. over 19 years), with 600 patients at risk in the external data at 6-years post-baseline. This was reflective of the SEER population registry (see Case Study). External data was generated from the same underlying data generating mechanism as the control arm of the trial but with an additional bias parameter to represent deviations in the hazard rates between external and trial patients. This bias might in practice be attributable to unobserved confounding, information biases in real-world sources[14], and further differences between trial and real-world populations (even after population-adjustment methods[19] have been applied). We considered scenarios where the external data had an all-cause hazard with relative bias of -20% to +20%. We simulated all survival data using the cumulative hazard inversion method[21] in the R software, using numerical integration to evaluate the cumulative hazard[21].

**Bayesian survival models**

We fitted Bayesian flexible parametric models to each simulated dataset using the `survextrap` package (version 0.8.12) [15]. We considered models that both incorporated and omitted the external data. When incorporating both trial and external datasets jointly in the model, we assumed the external data have the same survival as the control arm of the trial. Following the recommended defaults[13], we set the M-spline degrees of freedom (df) to be the default df = 10, with knots specified at quantiles of event times from the trial data, and we set a vague $\sigma$ smoothness prior of Gamma(2,1), and a weighted random walk prior on the spline coefficients (details in Supplementary Material). For the position of long-term knots, we compare the impact of alternative choices: 1) no additional knots beyond trial follow-up, 2) one additional at 25 years, 2) two additional knots at 10 and 25 years, and 3) three additional knots at 5, 10 and 25 years. All models fitted included background mortality rates, which were based on the true age-specific rates from the Gompertz distribution.

To represent treatment effects, we performed three analyses, one in which the control and active arms are modelled separately, one using a PH model, and one using a Non-PH model (details in Supplementary Material). As in the case study, these are used in combination with different assumptions about whether and how quickly the treatment effect wanes in the long term.

**Estimands and Performance Measures**

Our primary estimand was the marginal restricted mean survival time (RMST) at 40 years for the control arm and the difference in marginal RMST at 40 years (RMSTD) for the treatment effect between active and control arms, using the posterior median with 95% credible intervals. Performance measures were the bias, mean squared error (MSE), model posterior standard deviation (model SD), and coverage of the credible intervals.

# Results

## Case Study

**Within-trial model fit to the control arm**

We find that using 10 degrees of freedom (df) and a Gamma(2,1) prior placed on the smoothness parameter $\sigma$ is a sensible default model choice, providing good fit within the 5-year trial period, and hence these settings are used for all remaining models (Figure 2). Within-trial fit in the control arm is broadly insensitive to various model choices considered (Supplementary Table 1 and Supplementary Figure 1).

*Knot placements after end of trial follow-up*

When we include extra knots to account for structural uncertainty in a model without external data (Figure 2a), we note the high uncertainty in the extrapolated period suggests there is not enough information in the trial data alone to reliably estimate the long-term survival. In contrast, if (naively) no knots are placed outside the trial data, then the extrapolated hazard is constant with a narrow credible interval, and structural uncertainty about the long-term shape is not represented (Supplementary Figure 2). The beliefs about the long-term hazard that these extra knots represent can be illustrated by sampling hazard trajectories from the posterior distribution of hazards given the trial data only (Supplementary Figure 3). Note the prior on the smoothness parameter $\sigma$ prevents the curves from becoming too "wiggly" as more knots are added[6].

*Modelling the external data*

The use of population rates in an excess hazards model causes the projected long-term all-cause hazard to increase proportionally to the population rates (Figure 2b). This in turn reduces the uncertainty on the extrapolated survival, ensuring the survival function reaches zero by the lifetime horizon. When additionally incorporating SEER data the long-term hazard and survival estimates are precise, and relatively unaffected by knot placement (Supplementary Figure 2) indicating the registry data provide substantial information.

*Modelling the treatment effect*

Firstly we illustrate the models where no treatment effect waning is imposed. PH, Non-PH and separate arm modelling give similar estimates of the short-term treatment effect from the trial data (Figure 3). The PH model resulted in narrow credible intervals for the extrapolated hazard ratio (Figure 3a). By contrast, for Non-PH and separate arm models, the extrapolated active arm hazards, hence the hazard ratios, are extremely uncertain (Figure 3b and 3c) since the external registry data only describe the control group. Note the Non-PH model borrows a small amount of information from the control group (via some shared spline parameters, see Supplementary Methods) to estimate the treatment group hazard, giving a slightly more precise hazard ratio estimate than the separate arms model.

Modelling in an excess hazard framework caused the hazard ratio (describing all causes of death) to wane over time, and credible intervals to narrow. This occurs because the all-cause hazard becomes dominated by the general population rates in both arms over time (Figure 3, Supplementary Figure 4).

Unless PH is assumed, the extent of uncertainty about the treatment effect is also sensitive to the exact position of the external knots (Supplementary Figure 4) given that there is no long-term information about the treatment effect.

A lifetime treatment benefit is predicted from all the models, though there is considerable uncertainty in the magnitude of the difference in 40-year RMST (Table 1). This is the consequence of the different assumptions about how the treatment effect changes in the long term, which involve naively extrapolating from the short-term data in different ways. Given the lack of long-term evidence on the treatment effect, sensitivity analyses to this effect are needed, and this is done here with the treatment effect waning models. Under these models, the credible intervals about the treatment effect become narrower as the waning rate increases (Table 1).

## Simulation Study

### Control arm

*Analyses without external data.*

Survival and treatment effect data generating mechanisms for the simulation study are shown in Figure 4. A model with only trial data and population rates and with no extra knots overestimates the hazard at the end of trial follow-up, and this persists resulting in pessimistic extrapolations (Figure 5b) and hence considerable bias in the control arm RMST at 40-years (Figure 6a). To relax the constant excess hazard assumption and better account for structural uncertainty, knots outside the trial data, at 5, 10 and 25 years, are added, allowing the hazard to vary in the long-term. However, this only partially alleviates the bias (Figure 5b, Figure 6a), and the credible intervals are underconfident (Table 2). A model of this kind is still helpful, however, to demonstrate the need for long-term data, rather than as a way of accurately estimating the long-term hazard.

*The impact of including external data with/without bias.*

When including external data we find that estimates of 40-year RMST for the control arm are somewhat sensitive to the level of bias present in the external data (Figure 6a). This shows that the quality and relevance of longer-term external data, or how it is modelled, will be a factor that impacts extrapolation accuracy. However, scenarios with high levels of bias still outperform models where external data are omitted.

*Trial data cut-off*

Supplementary Figure 5 shows results from analyses that use trial data based on a cut-off at 3 years (44% survival in control arm), 5 years (34% survival) or 8 years (26% survival). Without external data, a trial cut-off of 8-years was required to provide unbiased extrapolations. With a trial data cut-off of 3-years, models that included unbiased external data provided unbiased estimates of control arm 40-year RMST.

*Knot placement*

Estimates of extrapolated RMST are slightly improved by adding more knots (df) inside the trial follow-up, however the accuracy of the extrapolation is most affected by whether external data are included (Supplementary Figure 6a). Placing two extra knots, at 10 and 25 years, appears to be sufficient to characterize the changes in hazard observed in the external data.

**Modelling the treatment effect**

*External data and Model specification (PH, Non-PH and Separate arms with no treatment effect waning).*

In scenario 1 (generating data with constant excess hazard ratio) we find that by incorporating external data on the control arm and using a proportional hazards model, we can accurately identify the lifetime treatment benefit (RMSTD) (Figure 6b). Further, informing the control arm survival with biased external data did not substantially bias estimates of the treatment effect. In this scenario, the Non-PH model also provides unbiased estimates of the RMSTD, whether or not external data are used, since the extrapolation is based on projecting a hazard ratio function learned from the short-term data, which was roughly constant over time. However, we would not recommend this approach in general, unless there is evidence that the hazard ratio will be constant in the long term. The separate-arms model estimates the long-term treatment effect poorly, even with external data, since this model does not include any data or assumption about the active arm in the long term. Coverage of the credible intervals for the RMSTD were close to the nominal value for PH models with or without external data if extra knots were included to allow for structural uncertainty about the baseline hazard beyond trial follow-up (Table 3).

For simulated data with a waning treatment effect (scenario 2), we find that the Non-PH model resulted in the smallest bias for the 40-year RMSTD (Figure 6c). In this scenario the treatment effect waning fully occurred during the period of the trial and hence the Non-PH model was able to accurately model the waning effect. The PH model did not perform as well here, showing that care needs to be taken applying PH assumptions if there is a waning effect. Separate arm models performed poorly in this case, since they include no long-term information about the active arm. The coverage statistics were variable across the models fitted, with overly wider credible intervals using the Non-PH models.

For a delayed then waning effect (scenario 3), both the PH and Non-PH models overestimate the treatment effect, since the true hazard ratio continues to wane after the trial ends (from 5 to 8 years) (Figure 6d). Without longer term data the extrapolations will tend to expect some treatment benefit to persist, and thus overestimate the treatment effect. The separate-arms model with no external data appeared to estimate the long-term treatment effect accurately but this was a chance finding, since bias was present in both control and active arms that cancelled out (Supplementary Figure 7).

*Sensitivity analyses to the extent of treatment effect waning.*

We note that in both scenarios 2 (waning effect) and 3 (delayed effect then waning), the PH model does not describe the true variations in the treatment effect in the trial period. Despite this misspecification, adequate estimates can be obtained by imposing the waning assumptions that are closest to the truth (Supplementary Figure 8). For example, rapid waning over a period of 1-5 years gives the lowest bias for Scenario 2, and the more gradual waning over 10-20 years performs best for Scenario 3 (compare with the true treatment effect trajectories in Figure 4).

Note that including external data has some impact (albeit small) since it improves estimation of the control group hazard, which the active arm reverts to after the waning period.

As the Non-PH model accounts for the true treatment effect variations in the trial period, when waning assumptions are imposed, the model fits well in both the short and long term, and accurate estimates of the RMSTD are obtained. Note that with the faster waning present in scenario 2, the estimates were also robust to the assumed time when waning reaches a null effect.

# Discussion

In summary, we present a case study and simulation to assess the performance and guide the use of flexible Bayesian evidence synthesis models within the `survextrap` R package for performing survival extrapolation, with a focus on the inclusion of external real-world evidence and estimation of long-term survival and treatment effects.

First, we show that incorporating real-world evidence can substantially increase the confidence in extrapolations, while also demonstrating that unbiased estimation of long-term survival and treatment effects can be achieved provided that long-term external data is present, or that appropriate assumptions relating to the treatment effect hold true (such as proportional hazards or imposed waning constraints). Second, we demonstrate the robustness of `survextrap` when modelling external data that was imperfect (e.g., that might arise from unobserved confounding, population differences, and other selection and information biases present in real-world sources[14]), as reflected by having relatively higher or lower hazard rates than the truth. We found that incorporating this data still improved the quality of extrapolations in comparison to relying on trial data alone. Third, we note `survextrap` has a degree of robustness to the "nuisance" choices inherent in selecting a model for extrapolation, such as the number and placement of knots and the prior distributions that govern hazard flexibility across the time horizon.

This work builds on findings from our previous simulation study that focused on showing `survextrap` models provide good fit to trial data in the short-term[13], as well as other studies that have focused on extrapolating single trial arms using only trial data with standard and more flexible parametric models[5,22,15,23]. A previous study by Vickers[8] performed simulations to assess the use of Bayesian flexible spline-based models when incorporating both short-term trial data and longer-term external data in aggregate count format, though this approach differed from ours in that the external data was first matched to the trial data using acceleration factor adjustments of the survival times, while the model had limited flexibility with only one or two knots, and stronger implicit treatment waning assumptions were made through the specification of priors. The Vickers study encountered challenges in both convergence and model fit in their JAGS[24] implementation, and concluded the findings had limited generalisability. Our work builds on this with a more comprehensive and transparent assessment of modelling assumptions, while demonstrating good fit even with more complex models.

Additionally, our work highlights the value of `survextrap` as a means to better express the 'structural uncertainty'[25,26] inherent in the shape of the hazard and survival curves. The M-spline hazard curve is arbitrarily flexible in the period spanned by its knots. Therefore by fitting `survextrap` models with knots placed beyond the duration of trial follow-up, the posterior from a model with no long-term data essentially acts as a "prior" for long-term survival, that assumes very little on how the hazard plausibly varies in the long-term. This reflects structural uncertainty about the hazard function due to lack of data, and demonstrates the potential value of more data to inform decision-making (which might be formally quantified with Value of Information methods[27–29]).

Note that the models we demonstrate, and our evaluations of these models, have some limitations. First, for our case study we were limited by lacking individual patient data (IPD) on the clinical trial and so we used population mortality rates based on mean age at diagnosis for all individuals as a simplistic assumption. In practise, we would recommend investigating variation of survival by covariates, especially age, and deriving marginal estimates for decision making through regression standardisation[30]. Second, our simulation study depends on a particular data generating mechanism from a single clinical trial. Future work could consider different design choices such as varying the sample size of the trial and external data, cure model scenarios[31,32], and the extent to which the follow-up period of trial and external data overlaps. Future research may also wish to explore the use of `survextrap` to control the relative weight and contribution of the external data in cases where it is suspected to be biased[10], such as by extrapolating assuming proportional hazards between dataset populations, and through the use of power priors or power likelihoods[33,34], or hierarchical priors[33].

To conclude, our work highlights the potential value of including longer-term real-world evidence to anchor survival extrapolations using a flexible Bayesian approach, and helps to guide practitioners on applying modelling assumptions and sensitivity analyses to evaluate longer-term treatment effects. These findings help provide greater confidence in the use and application of this class of survival models for both modellers and decision-makers.

**Figure 1:** Illustrative example of `survextrap` models fitted to the radiotherapy control arm from Bonner et al trial without inclusion of external data. a) model with df = 10 with no extra knots beyond duration of trial, resulting in a constant hazard extrapolation from the end of the trial, b) model with df = 10 with extra knots at 10, 15 and 25 years, which enables the hazard to vary beyond the trial period, capturing the 'structural uncertainty' in the extrapolated period. Knot locations in both examples are shown with the solid vertical lines. Dashed vertical line represent the end of trial follow-up period.

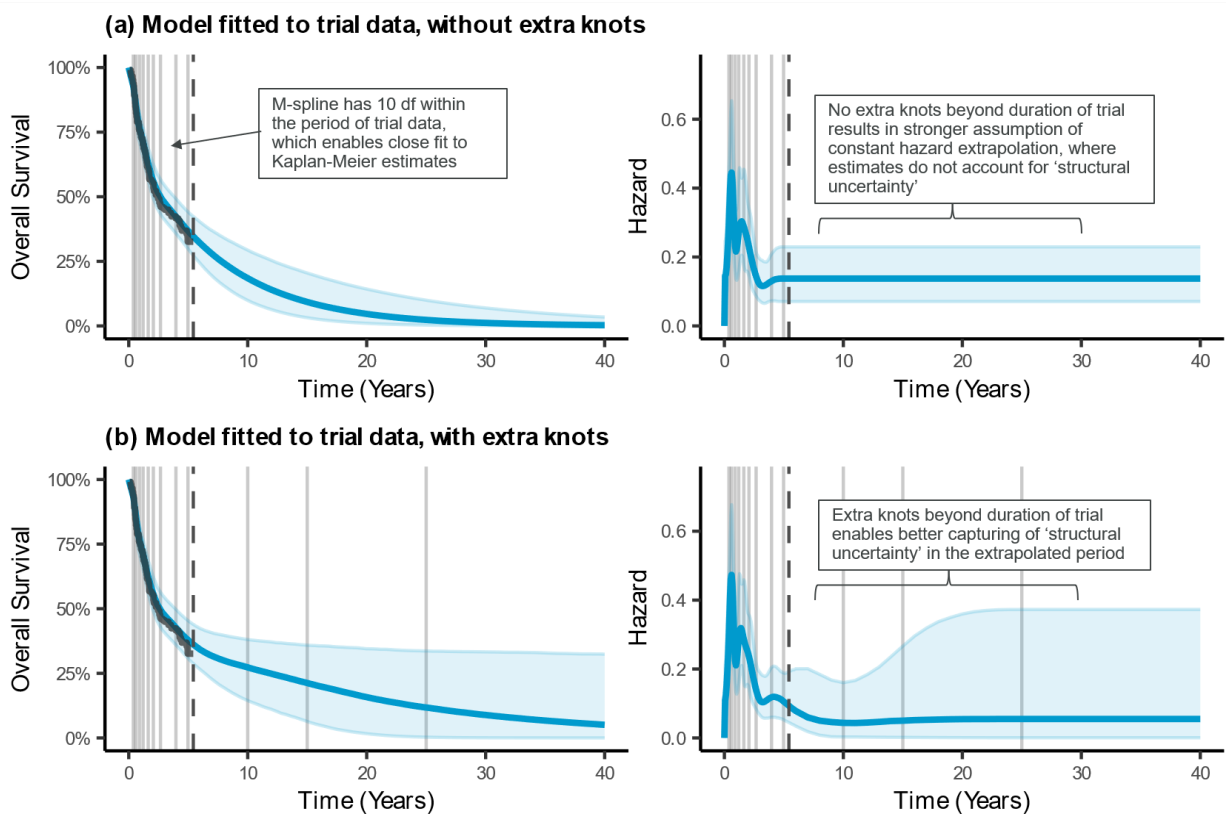

**Figure 2:** Survival extrapolations for the radiotherapy arm of the Bonner et al trial, with and without external data (population mortality rates and SEER registry data). Dashed vertical line represents the end of trial follow-up period. Each model uses 10 df within the trial duration with extra knots placed at 10, 15 and 25 years. Posterior medians and 95% credible intervals.

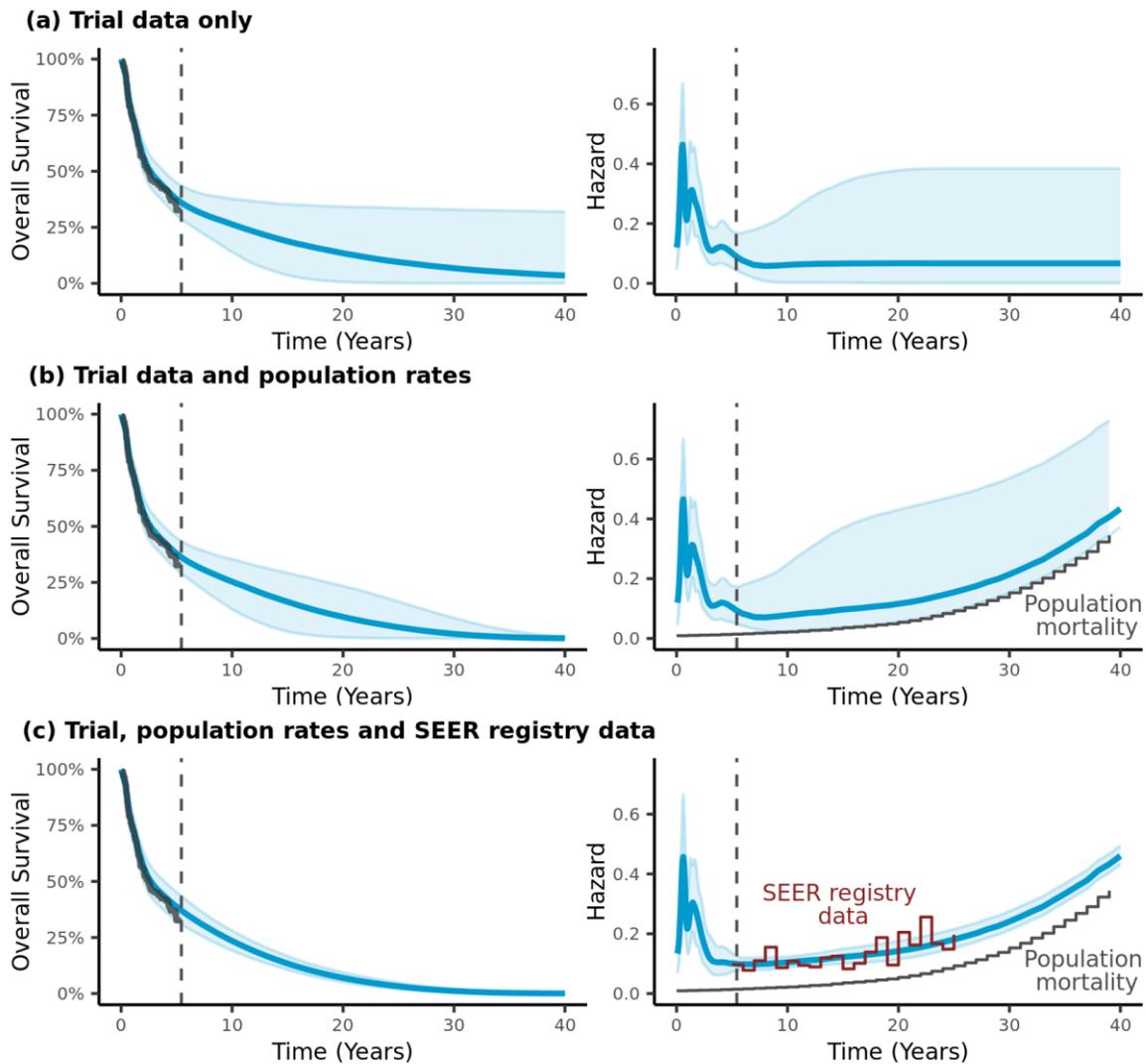

**Figure 3:** Survival extrapolations of both arms of the Bonner et al. trial. All models incorporate clinical trial data, population rates and SEER registry data, fitted using (a) proportional excess hazard models, (b) flexible non-proportional hazards models, and (c) separate arms modelling. The vertical line indicates the end of the trial follow-up period. Each model uses 10 df within trial duration with extra knots placed at 10, 15 and 25 years. Posterior medians and 95% credible intervals.

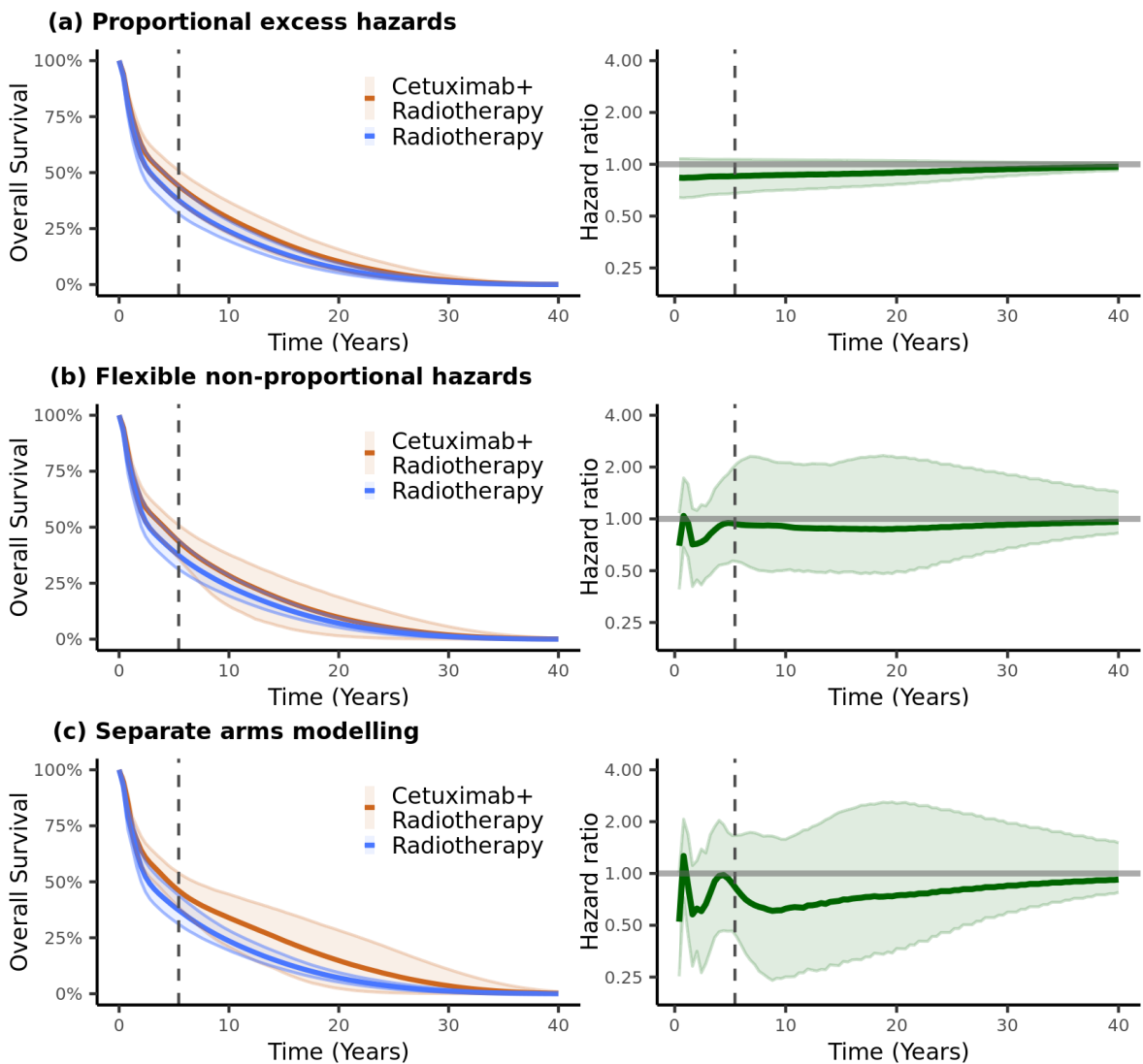

**Figure 4:** True marginal survival and hazard / hazard ratio functions for the simulation study. (a) Survival and hazard functions for the control arm, with radiotherapy Kaplan-Meier curve from Bonner trial superimposed alongside survival, and general population mortality (GPM) rates alongside the hazard, respectively. (b), (c), and (d) Survival for control and active arms and hazard ratio function based on (b) constant excess hazard ratio, (c) waning excess hazard ratio, (d) delayed then waning excess hazard ratio. The 5-y end of trial data cut-off has been shown as a vertical dashed line.

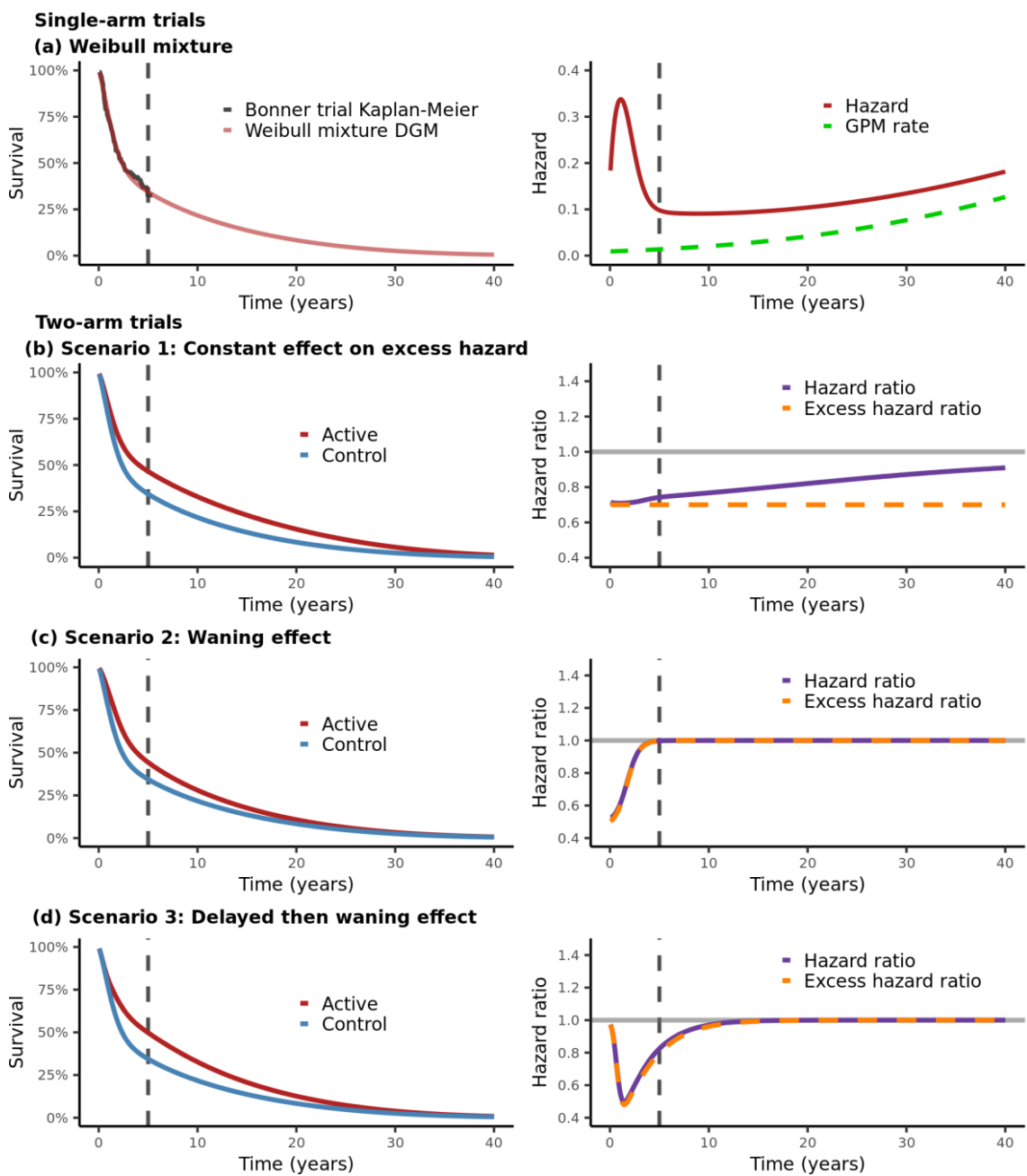

**Figure 5:** Estimated marginal survival and hazard curves for `survextrap` models fitted to 25 simulated control arm datasets. The true marginal survival and hazard function is shown as a solid black line. The top row of each panel shows models fitted without external data without or with extra knots. The bottom row of each panel shows a model with extra knots fitted to external data with varying levels of bias.

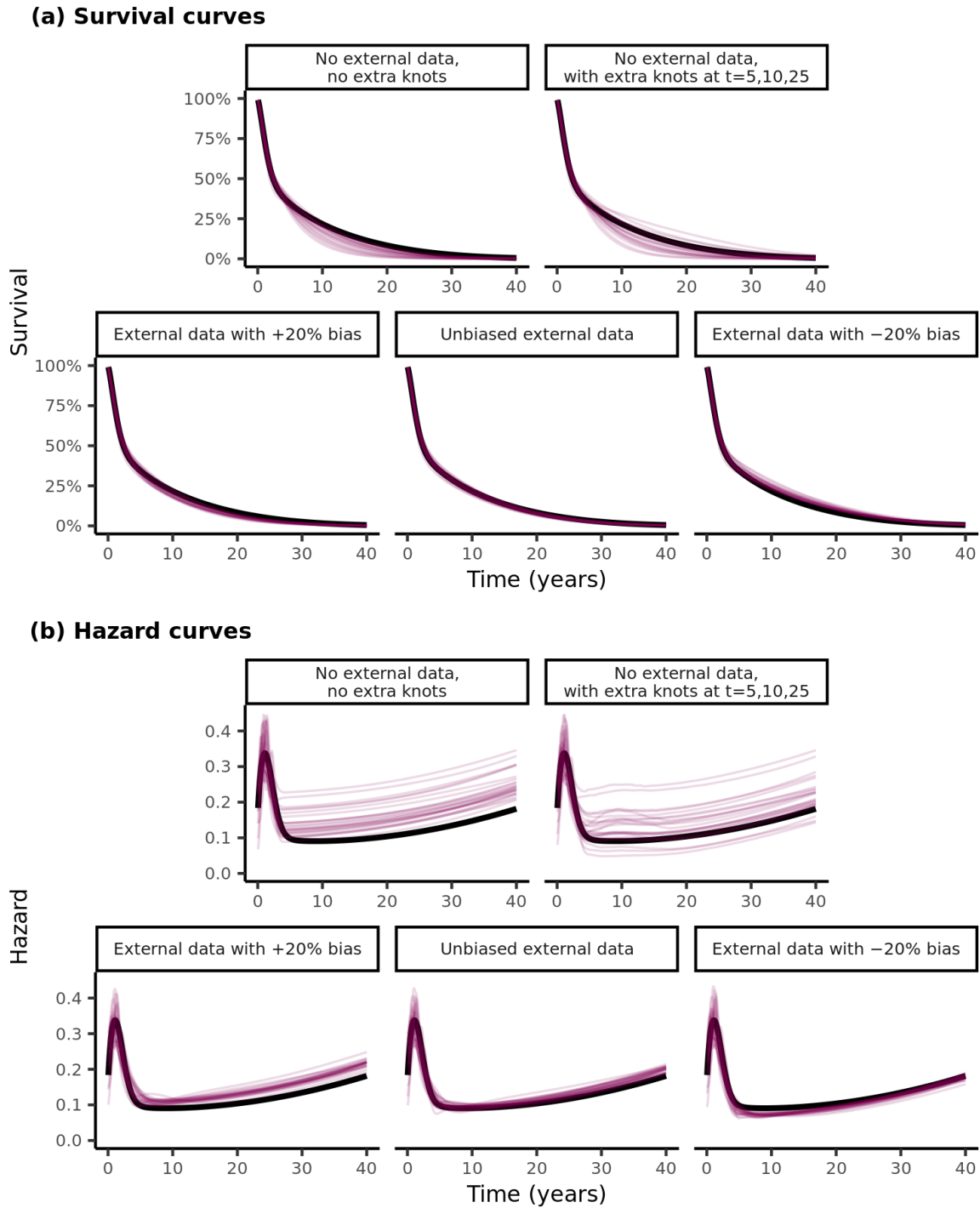

**Figure 6:** Bias in (a) marginal 40-year RMST in the control arm, (b), (c), (d) marginal 40-year RMST difference between active and control arms, under various data generating mechanisms (scenarios, external data bias) and model choices (with / without external data, with / without extra knots, PH, Non-PH and separate arm treatment effect models). The vertical line shows the true value.

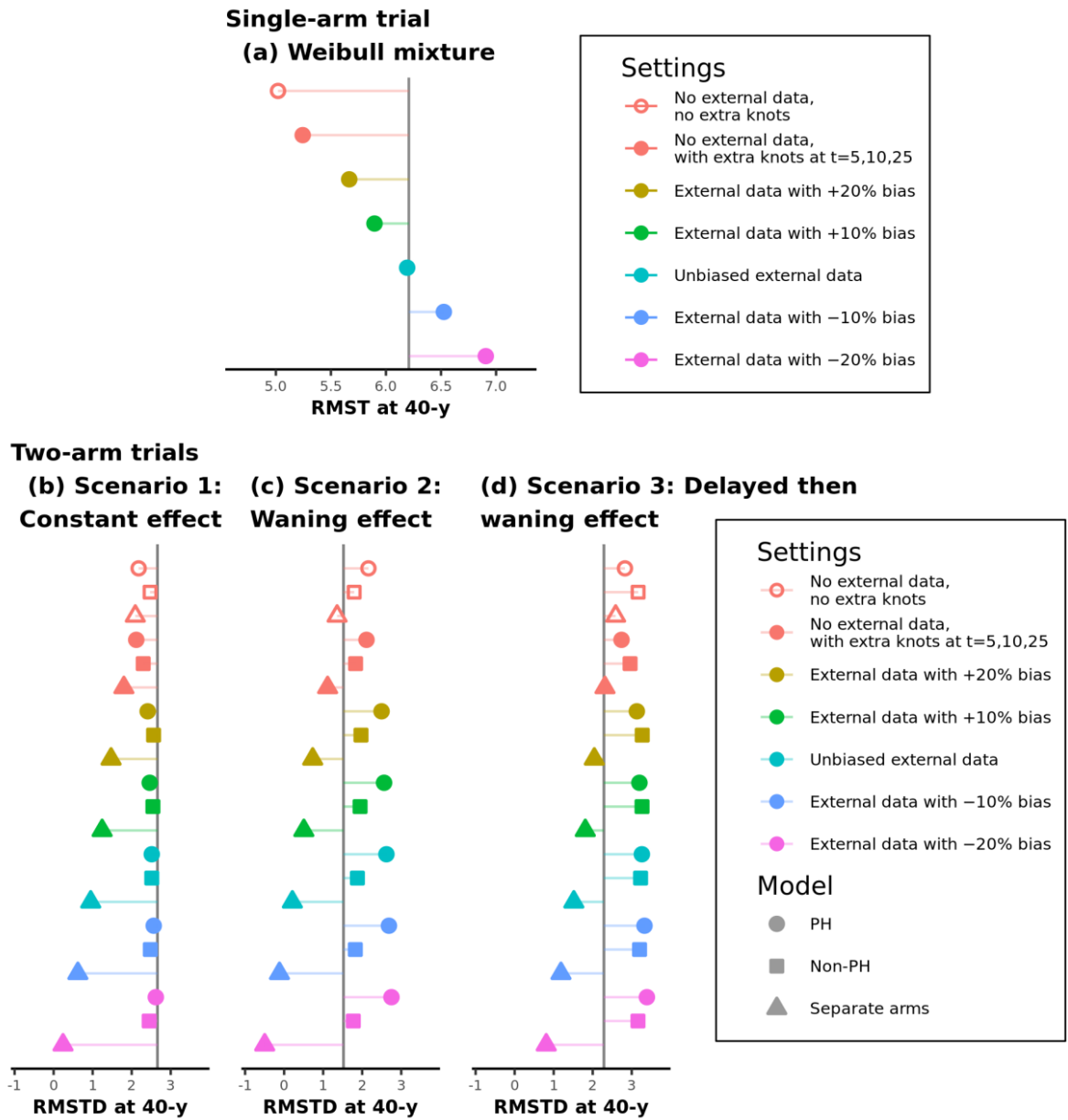

**Table 1:** Estimates of restricted mean survival time (RMST) and difference in RMST (RMSTD) at 40-years for radiotherapy and cetuximab plus radiotherapy arms. All models presented use an M-spline with 10 df within the trial period, and extra knots placed at 10, 15 and 25 years.

| Model | Waning time (years) | Control arm, RMST at 40-y (95% credible interval) | | | Difference in RMST at 40-y (95% credible interval) | | |
|---|---|---|---|---|---|---|---|
| | | Trial only | Trial + Population rates | Trial + Population rates + Registry | Trial only | Trial + Population rates | Trial + Population rates + Registry |
| Proportional hazards | - | 7.27 (4.52, 13.40) | 6.53 (4.51, 9.08) | 6.32 (5.45, 7.19) | 1.97 (-0.43, 5.60) | 1.43 (-0.42, 3.43) | 1.23 (-0.46, 3.12) |
| Flexible non-proportional hazards | - | 7.12 (4.64, 13.47) | 6.41 (4.59, 9.11) | 6.29 (5.41, 7.20) | 1.73 (-2.01, 5.96) | 1.32 (-1.00, 3.75) | 1.09 (-1.13, 3.43) |
| Separate arms | - | 8.22 (4.76, 15.07) | 6.94 (4.71, 9.71) | 6.27 (5.43, 7.21) | 1.77 (-6.32, 10.16) | 1.55 (-2.57, 5.64) | 2.27 (-0.46, 5.37) |
| Proportional hazards | 6 | 7.27 (4.52, 13.40) | 6.53 (4.51, 9.08) | 6.32 (5.45, 7.19) | 1.25 (-0.29, 3.60) | 1.04 (-0.33, 2.56) | 0.84 (-0.33, 2.07) |
| Proportional hazards | 10 | 7.27 (4.52, 13.40) | 6.53 (4.51, 9.08) | 6.32 (5.45, 7.19) | 1.41 (-0.33, 4.08) | 1.15 (-0.36, 2.85) | 0.93 (-0.37, 2.32) |
| Proportional hazards | 20 | 7.27 (4.52, 13.40) | 6.53 (4.51, 9.08) | 6.32 (5.45, 7.19) | 1.61 (-0.36, 4.63) | 1.28 (-0.39, 3.10) | 1.08 (-0.42, 2.68) |

**Table 2:** Performance measures with Monte Carlo standard errors in parentheses for estimates of control arm 40-year RMST. Extra knots are placed at $t$ = 5, 10 and 25 years.

| Settings | True value | Bias | MSE | Model SD | Coverage |
|---|---|---|---|---|---|
| No external data, no extra knots | 6.21 | -1.19 (0.026) | 2.06 (0.054) | 0.94 (0.009) | 0.77 (0.013) |
| No external data, with extra knots | 6.21 | -0.96 (0.032) | 1.98 (0.059) | 1.76 (0.011) | 0.99 (0.003) |
| External data with +20% bias | 6.21 | -0.54 (0.013) | 0.47 (0.016) | 0.40 (0.001) | 0.74 (0.014) |
| External data with +10% bias | 6.21 | -0.31 (0.014) | 0.28 (0.011) | 0.43 (0.001) | 0.88 (0.010) |
| Unbiased external data | 6.21 | -0.02 (0.015) | 0.22 (0.010) | 0.45 (0.001) | 0.94 (0.008) |
| External data with -10% bias | 6.21 | 0.32 (0.016) | 0.34 (0.016) | 0.48 (0.001) | 0.90 (0.010) |
| External data with -20% bias | 6.21 | 0.70 (0.016) | 0.76 (0.027) | 0.52 (0.001) | 0.72 (0.014) |

**Table 3:** Performance measures with Monte Carlo standard errors in parentheses for estimates of the difference in RMST at 40-years for active versus control arms. Extra knots are placed at t = 5, 10 and 25 years.

| Settings | Model | True value | Bias | MSE | Model SD | Coverage |
|---|---|---|---|---|---|---|
| *Scenario 1: Constant effect* | | | | | | |
| No external data, no extra knots | PH | 2.66 | -0.49 (0.030) | 1.11 (0.047) | 0.93 (0.003) | 0.90 (0.009) |
| No external data, no extra knots | Non-PH | 2.66 | -0.19 (0.036) | 1.32 (0.060) | 1.39 (0.007) | 0.98 (0.005) |
| No external data, no extra knots | Separate arms | 2.66 | -0.57 (0.042) | 2.09 (0.093) | 1.61 (0.009) | 0.96 (0.006) |
| No external data, with extra knots | PH | 2.66 | -0.55 (0.030) | 1.19 (0.049) | 1.14 (0.005) | 0.95 (0.007) |
| No external data, with extra knots | Non-PH | 2.66 | -0.37 (0.033) | 1.24 (0.051) | 1.90 (0.008) | 1.00 (0.001) |
| No external data, with extra knots | Separate arms | 2.66 | -0.86 (0.045) | 2.73 (0.123) | 2.20 (0.011) | 0.98 (0.004) |
| Unbiased external data | PH | 2.66 | -0.15 (0.035) | 1.22 (0.054) | 1.07 (0.002) | 0.94 (0.008) |
| Unbiased external data | Non-PH | 2.66 | -0.15 (0.037) | 1.43 (0.064) | 1.99 (0.007) | 1.00 (0.001) |
| Unbiased external data | Separate arms | 2.66 | -1.71 (0.038) | 4.35 (0.120) | 1.39 (0.009) | 0.80 (0.013) |
| *Scenario 2: Waning effect* | | | | | | |
| No external data, no extra knots | PH | 1.52 | 0.64 (0.027) | 1.14 (0.051) | 0.88 (0.003) | 0.89 (0.010) |
| No external data, no extra knots | Non-PH | 1.52 | 0.28 (0.033) | 1.19 (0.057) | 1.30 (0.007) | 0.97 (0.005) |
| No external data, no extra knots | Separate arms | 1.52 | -0.16 (0.038) | 1.49 (0.074) | 1.47 (0.009) | 0.97 (0.006) |
| No external data, with extra knots | PH | 1.52 | 0.59 (0.028) | 1.11 (0.052) | 1.12 (0.006) | 0.96 (0.006) |
| No external data, with extra knots | Non-PH | 1.52 | 0.31 (0.031) | 1.06 (0.051) | 1.88 (0.009) | 1.00 (0.001) |
| No external data, with extra knots | Separate arms | 1.52 | -0.40 (0.042) | 1.94 (0.103) | 2.09 (0.011) | 0.99 (0.003) |
| Unbiased external data | PH | 1.52 | 1.10 (0.034) | 2.37 (0.092) | 1.07 (0.002) | 0.83 (0.012) |
| Unbiased external data | Non-PH | 1.52 | 0.36 (0.039) | 1.66 (0.070) | 2.04 (0.007) | 1.00 (0.000) |
| Unbiased external data | Separate arms | 1.52 | -1.31 (0.033) | 2.79 (0.086) | 1.23 (0.009) | 0.83 (0.012) |
| *Scenario 3: Delayed then waning effect* | | | | | | |
| No external data, no extra knots | PH | 2.28 | 0.54 (0.031) | 1.24 (0.059) | 0.98 (0.003) | 0.90 (0.009) |
| No external data, no extra knots | Non-PH | 2.28 | 0.88 (0.040) | 2.35 (0.113) | 1.48 (0.007) | 0.93 (0.008) |
| No external data, no extra knots | Separate arms | 2.28 | 0.30 (0.045) | 2.09 (0.111) | 1.67 (0.009) | 0.96 (0.006) |
| No external data, with extra knots | PH | 2.28 | 0.45 (0.031) | 1.19 (0.059) | 1.23 (0.006) | 0.99 (0.004) |
| No external data, with extra knots | Non-PH | 2.28 | 0.67 (0.037) | 1.80 (0.089) | 2.12 (0.009) | 1.00 (0.001) |
| No external data, with extra knots | Separate arms | 2.28 | 0.03 (0.047) | 2.18 (0.114) | 2.23 (0.011) | 0.99 (0.003) |
| Unbiased external data | PH | 2.28 | 0.98 (0.036) | 2.26 (0.099) | 1.12 (0.002) | 0.86 (0.011) |
| Unbiased external data | Non-PH | 2.28 | 0.94 (0.042) | 2.69 (0.116) | 2.22 (0.009) | 1.00 (0.001) |
| Unbiased external data | Separate arms | 2.28 | -0.77 (0.040) | 2.15 (0.085) | 1.43 (0.009) | 0.93 (0.008) |

# Competing Interests

The authors declared the following potential conflicts of interest with respect to the research, authorship, and/or publication of this article: IRT and MJS are full-time employees of AstraZeneca. MJS reports AstraZeneca stock ownership. IRT is a fellow of the AstraZeneca Postdoctoral Research Programme.

The authors disclosed receipt of the following financial support for the research, authorship, and/or publication of this article: Financial support for this study was provided in part by a contract with AstraZeneca. The funding agreement ensured the authors' independence in designing the study, interpreting the data, writing, and publishing the report. The following authors are employed by the sponsor: IRT, MJS.

CJ is funded by the Medical Research Council, programme code MC_UU_00040/4. FT is funded by the UKRI-MRC - programme number MR/T033371/1.

# Data Availability

All data and code that support the findings of this study are available at the following URL: https://github.com/irtimmins/Simulation-study2-RWE-survextrap

# Supplementary Methods

## Details of the `survextrap` model

### An M-spline for the hazard function

The model uses M-splines to describe the hazard function (or the excess hazard function if background mortality rates are included), which is defined as a weighted sum of $n$ basis functions $b_i(t)$ of the form:

$$h(t) = \eta \sum_{i=1}^{n} p_i\, b_i(t)$$

with scale parameter $\eta$ and basis coefficients $p_i$ where $\sum_i p_i = 1$. The basis functions are polynomials (cubic by default), restricted to be positive, with continuity constraints at the knot locations such that the resulting hazard is a smooth, positive-valued function. The flexibility of the spline is determined by the number of basis functions and the knot locations. The knots are, by default, placed at equally-spaced quantiles of the event times in the trial. We previously investigated the impact of varying the number of knots when modelling the trial data alone and found results are relatively insensitive to this choice[1].

The Bayesian model can be extended to allow the hazard to depend on explanatory variables such as treatment; and the effect of these can either be constant through time with the assumption of proportional hazards, or time-varying (i.e. a non-proportional hazards model). In the proportional hazards (PH) model, the scale parameter is redefined with covariates $\boldsymbol{\beta}$ and explanatory variables **x**:

$$\eta(\mathbf{x}) = \eta_0 \exp(\boldsymbol{\beta}^T \mathbf{x})$$

In addition, a flexible non-proportional hazards (non-PH) model allows explanatory variables to affect the M-spline coefficients through a multinomial logistic regression where $\log(p_i(\mathbf{x})/p_1(\mathbf{x})) = \gamma_i(\mathbf{x}) = \mu_i + \boldsymbol{\delta}_i^T \mathbf{x} + \sigma\epsilon_i$. The $s$th element of the vector $\boldsymbol{\delta}_i$ describes the amount of departure from a proportional hazards model for the $s$th covariate in the region of time associated with the $i$th basis term. Hence, if $\delta_{is} = 0$ for all $i$, then the $s$th covariate follows proportional hazards. A hierarchical prior is used for these coefficients such that $\delta_{is} \sim N(0, \tau_s)$, which smooths the covariate effects over the time regions. A relatively weak Gamma(2,1) non-informative prior is the default option for each of the $\tau_s$.

**Prior specification**

Priors, potentially informative, are placed on the scale parameter and each of the basis coefficients. A hierarchical random walk prior on the basis coefficients controls the smoothness of the spline (available from `survextrap` v0.8.16). Specifically the coefficients are modelled as $\log\left(\frac{p_i}{p_1}\right) = \mu_i + \sigma\epsilon_i$ and $\epsilon_i \sim Logistic(\epsilon_{i-1}, w_i)$, where $\sigma$ controls the magnitude of deviation from a constant hazard and $w_i$ is a constant that accounts for the difference in the time periods described by coefficients $i$ and $i-1$ (see Phillippo et al.[2]).

**Modelling on the excess hazard scale**

The model also supports relative survival models (also known as excess hazard models). In this framework, the overall hazard $h(t) = h^*(t) + \gamma(t)$ is the sum of the background mortality rate $h^*(t)$ and excess hazard $\gamma(t)$. The excess hazard $\gamma(t)$ is modelled by the M-spline and covariates. The background mortality rate is treated like an offset and is assumed fixed and known and is typically based on a lifetable matched by age, sex and calendar year.

**Stan model fit**

Models in `survextrap` are fitted using `rstan` which by default performs Hamiltonian Monte Carlo sampling from the posterior distribution of the model. Alternatively, a fast Laplace approximation to the full posterior distribution can be fitted, based on estimating the posterior mode by optimisation. In all analyses presented we used full Hamiltonian Monte Carlo sampling. We previously investigated the accuracy of the Laplace approximation, details of which can be found in[1].

**Model predictions**

Once the model has been fitted, predictions of the all-cause survival and the restricted mean survival time (RMST) can be obtained from the posterior samples. For excess hazard models the all-cause survival is obtained by combining relative survival predictions with background expected survival. The RMST is then the integral of the all-cause survival function up to a specific time point. If covariates are included in the model or in the background rates, marginal estimates of the all-cause survival and RMST are obtained via standardization.

**Marginal estimates through standardization**

When survival depends on covariate effects, marginal estimates of the survival, hazard and RMST distribution for each trial arm can be calculated via standardization. For $N$ patients with individual survival $S_i(t)$ for individual $i$, the marginal all-cause survival $\bar{S}(t)$ for this sample is given by

$$\bar{S}(t) = \frac{1}{N} \sum_{i=1}^{N} S_i(t)$$

Additionally, the hazard function for the marginal all-cause survival, where the hazard $h_i(t)$ denotes the hazard for individual $i$, is

$$\bar{h}(t) = \frac{\sum_{i=1}^{N} S_i(t)h_i(t)}{\sum_{i=1}^{N} S_i(t)}$$

Finally, the marginal restricted mean survival $\overline{RMST}(t)$, where restricted mean survival $RMST_i(t) = \int_0^t S_i(x)dx$ denotes the restricted mean survival for individual $i$ at time $t$, is

$$\overline{RMST}(t) = \frac{1}{N}\sum_{i=1}^{N} RMST_i(t)$$

**Treatment effect waning**

Treatment effect waning analyses can performed in `survextrap`. This requires us to first fit a parametric model $h(t|x)$ with treatment covariate $x = 0$ and $x = 1$ for the control and active arms, respectively. The predicted hazard for the control is given by $h(t|x = 0)$. Assume the treatment effect wanes over a time interval $t_{min}$ to $t_{max}$. Between time $t = 0$ and $t = t_{min}$, the predicted hazard in the active group is equal to the model estimate $h(t|x = 1)$. For the interval between $t = t_{min}$ and $t = t_{max}$, the log hazard ratio is assumed to wane linearly, which means the log hazard in the active group is given by

$$\log h(t|x = 0) + w(t) \log\left(\frac{h(t_{min}|x = 1)}{h(t_{min}|x = 0)}\right)$$

where $w(t) = (t_{max} - t)/(t_{max} - t_{min})$, and exponentiating this expression provides the hazard function for the active group. For times greater than $t = t_{max}$, the hazard in the active group is equal to the model estimate for the control of $h(t|x = 0)$.

## Prior and model choice for the case study

We investigated the fit of the Bayesian survextrap model to the control arm trial data. We specified an M-spline with either 3, 6 or 10 degrees of freedom (df) for modelling the individual-level patient data over the duration of the trial period with the last knot placed at the last OS event time. We specified for all models a prior for the scale parameter $\eta$ of $\log(\eta) \sim N(-2.73, 0.71)$, representing a mean survival of 25 years with a wide 95% credible interval of 5 to 100 years. We investigated three possible choices of prior for the smoothness parameter $\sigma$; ranging from very little restriction on the shape of the hazard function over time [Gamma(2,1)], to medium information [Gamma(2,5)], to a more restrictive, and strong, prior [Gamma(2,20)] (further information of these priors in[3]).

Leave-one out cross validation information criteria (LOOIC) and the fitted hazard function are very similar between these models (Supplementary Figure 1 and Supplementary Table 1).

When modelling the treatment effect a weak prior was given to the (log) hazard ratio of $N(0, 2.5)$ designed only to rule out extremely implausible values, with a prior median of 1 and upper 95% credible interval of 40 for the hazard ratio[3]. The hazard ratio variability parameter $\tau$ in Non-PH models was given a weak Gamma(2,1) prior.

## Simulation study data generating mechanisms

We defined a data-generating model for simulating survival times for 1,000 clinical trial and real-world datasets. For the trial datasets we considered a two-arm trial with $N = 400$ (200 per arm) with 1:1 randomisation, with follow-up of 5 years. The ages of trial participants were simulated from a normal distribution with mean 60 years and standard deviation 9 years. We modelled our control arm to fit the overall survival of patients with head and neck cancer treated with radiotherapy from the Bonner et al. trial [4], of which the Kaplan-Meier curve is shown Figure 1. The disease-specific (latent) survival was defined from a two-component Weibull mixture distribution. This parametric form was chosen as a flexible hazard that is mathematically different from a spline, to not overly favour the spline-based implementation of survextrap when fitting models to the data. The hazard function for the disease-specific survival is given by:

$$h_d(t) = h_{d0}(t) \exp(T\beta(t))$$

and cumulative hazard:

$$H_d(t) = \int_0^t h_{d0}(x) \exp(T\beta(x)) dx$$

where $T$ is a treatment covariate with $T = 0$ for the control arm and $T = 1$ for the active arm, and $\beta(t)$ is the (potentially time-varying) log hazard ratio for the effect of treatment, detailed in Supplementary Table 2. The baseline hazard is $h_{d0}(t) = \frac{p\, f_1(t) + (1-p) f_2(t)}{p\, S_1(t) + (1-p)\, S_2(t)}$, where $S_j(t) = \exp(-\lambda_j t^{\gamma_j})$ and $f_j(t) = \lambda_j \gamma_j t^{\gamma_j - 1} \exp(-\lambda_j t^{\gamma_j})$ are the survivor and density functions for the $j$th component of the mixture distribution, and $p$ is the mixture proportion. Based on the fit to the radiotherapy arm of the Bonner trial, the cause-specific mixture Weibull distribution had mixture probability $p = 0.41$ and individual Weibull components with shape and scale parameters $\gamma_1 = 1.53$ and $\lambda_1 = 0.52$, and $\gamma_2 = 0.82$ and $\lambda_1 = 0.13$, respectively.

For other-cause survival, we assumed a Gompertz distribution that was conditional on patients age $a$ at baseline:

$$S_{other}(t \mid a) = P(T \geq t + a \mid T \geq a) = S_{other}(t + a)/S_{other}(a)$$

where $S_{other}(t) = \exp\left(\lambda_{gpm}\, \gamma_{gpm}^{-1}\left(exp(\gamma_{gpm}\, t) - 1\right)\right)$. We set the Gompertz distribution to have rate and scale parameters $\lambda_{gpm} = 4.3 \times 10^{-5}$ and $\gamma_{gpm} = 9.4 \times 10^{-2}$, respectively, which were derived previously from English mortality rates by Rutherford et al. [5]. The all-cause survival was taken as the product of disease-specific and other-cause survival:

$$S(t|a) = S_d(t)\, S_{other}(t|a)$$

To generate data for the active arm we considered three treatment effect scenarios for the parameter $\beta(t)$: a constant disease-specific treatment effect giving rise to proportional excess hazards (scenario one), an immediate treatment effect followed by treatment waning (scenario two) and a delayed treatment effect followed by treatment waning (scenario three). The closed-form expressions for the hazard ratio functions in each scenario are shown in Supplementary Table 2 and can be seen in Figure 3. The marginal survival curves based on the three treatment effect scenarios are also shown in Figure 3.

We simulated survival data using the cumulative hazard inversion method[6] in R software. Since the cumulative hazard does not in general have a closed-form solution (unless the treatment effect $\beta(t)$ was constant, as in scenario one), we used numerical integration to evaluate $\int_0^t h_{d0}(x)\exp(T\beta(x))dx$ using Gauss-Legendre quadrature with 100 nodes.

We further assumed random censoring that would arise from staggered recruitment into the trial, by performing uniform censoring between 3 and 5 years.

In addition, external data was generated to inform control arm hazard rates from years 6 to 25 (i.e. over 19 years), with 600 patients at risk in the external data at 6-years post-baseline. This was reflective of the SEER population registry (see Case Study). We assumed cohort follow-up for the external data so that no new patients enter the risk set during the follow-up from year 6 to 25. External data was generated from the same underlying data generating mechanism as the control arm of the trial but with an additional bias parameter that represents deviation in the hazard rates between external and trial patients. Thus, the all-cause cumulative hazard function was scaled by a factor of $\exp(v)$, where $v$ is a bias parameter. We varied the bias parameter from between a relative bias of -20% to +20%, corresponding to $v = \log(0.8), \log(0.9), \log(1.0), \log(1.1)$ and $\log(1.2)$. There was no censoring distribution for external data patients.

True values of estimands were obtained by simulating one very large dataset (N=$10^8$) without censoring and empirically calculating the marginal RMST and marginal RMSTD using their sample mean. These were accurate to 2 decimal places based on Monte Carlo standard errors of the sample means.

## Prior and model choice for the simulation study

For the simulation study, we use for all models the default vague prior of $\log(\eta) \sim N(0, 20)$ on the scale parameter $\eta$. We also used a weak prior for the smoothness parameter $\sigma \sim \text{Gamma}(2,1)$, which imposed very little restriction on the shape. The treatment effect using PH models, we used a weak prior on the (log) hazard ratio of $N(0, 2.5)$. For Non-PH models, the hazard ratio variability parameter $\tau$ models was given a vague prior of $\text{Gamma}(2,3)$. All models used a random walk prior on the spline coefficients.

**Supplementary Table 1:** Goodness of fit for radiotherapy control arm from Bonner et al. trial, assessed with leave-one out cross validation information criterion (LOOIC) metrics.

| Spline degrees of freedom | Extra knots | Sigma prior | LOOIC | RMST at 5-y (95% credible interval) |
|---|---|---|---|---|
| 10 | - | Gamma(2, 1) | 592.6 | 2.90 (2.64, 3.15) |
| 3 | - | Gamma(2, 1) | 589.9 | 2.87 (2.62, 3.14) |
| 6 | - | Gamma(2, 1) | 595.1 | 2.89 (2.64, 3.14) |
| 10 | - | Gamma(2, 5) | 593.5 | 2.90 (2.63, 3.16) |
| 10 | - | Gamma(2, 20) | 595.5 | 2.91 (2.65, 3.17) |
| 10 | 25 | Gamma(2, 1) | 593.7 | 2.89 (2.63, 3.15) |
| 10 | 10,25 | Gamma(2, 1) | 593.3 | 2.89 (2.63, 3.17) |
| 10 | 10,15,25 | Gamma(2, 1) | 593.4 | 2.90 (2.64, 3.16) |

**Supplementary Table 2:** Treatment effect functions for each of the three simulation scenarios (see Figure 3 in main text for an illustration of these). These denote the ratio of disease specific hazard rates (excess hazard ratios).

| Scenario | Excess Hazard ratios comparing the active relative to control arm |
|---|---|
| 1: constant effect (proportional excess hazards) | $\exp[\beta(t)] = 0.7$ |
| 2: immediate effect and fast waning | $\exp[\beta(t)] = -0.38 + 0.38 \tanh(0.8t - 1.2)$<br><br>where $\tanh(x) = \frac{e^x + e^{-x}}{e^x - e^{-x}}$ is the hyperbolic tangent function. |
| 3: delayed effect and slower waning | $\exp[\beta(t)] = -2.8 \, f_{emg}(t; \mu = 0.8, \sigma = 0.4, \lambda = 0.35)$<br><br>where $f_{emg}$ is the probability density function of an exponentially modified Gaussian distribution:<br><br>$$f_{emg}(x; \mu, \sigma, \lambda) = \frac{\lambda}{2} e^{\frac{\lambda}{2}(2\mu + \lambda - 2x)} \text{erfc}\left(\frac{\mu + \lambda\sigma^2 - x}{\sqrt{2}\sigma}\right)$$<br><br>And $\text{erfc}$ is the complementary error function:<br><br>$$\text{erfc}(x) = 1 - \text{erf}(x) = \frac{2}{\sqrt{\pi}} \int_x^\infty e^{-s^2} \, ds$$ |

**Supplementary Figure 1:** Survival function (posterior median and 95% credible intervals) for the radiotherapy control arm from Bonner et al trial without inclusion of external data. Sensitivity analyses on a) varying the within-trial number of basis functions (degrees of freedom), b) the prior for sigma that controls the hazard smoothness, specified as Gamma(2, k) where k takes values 1, 5, or 20.

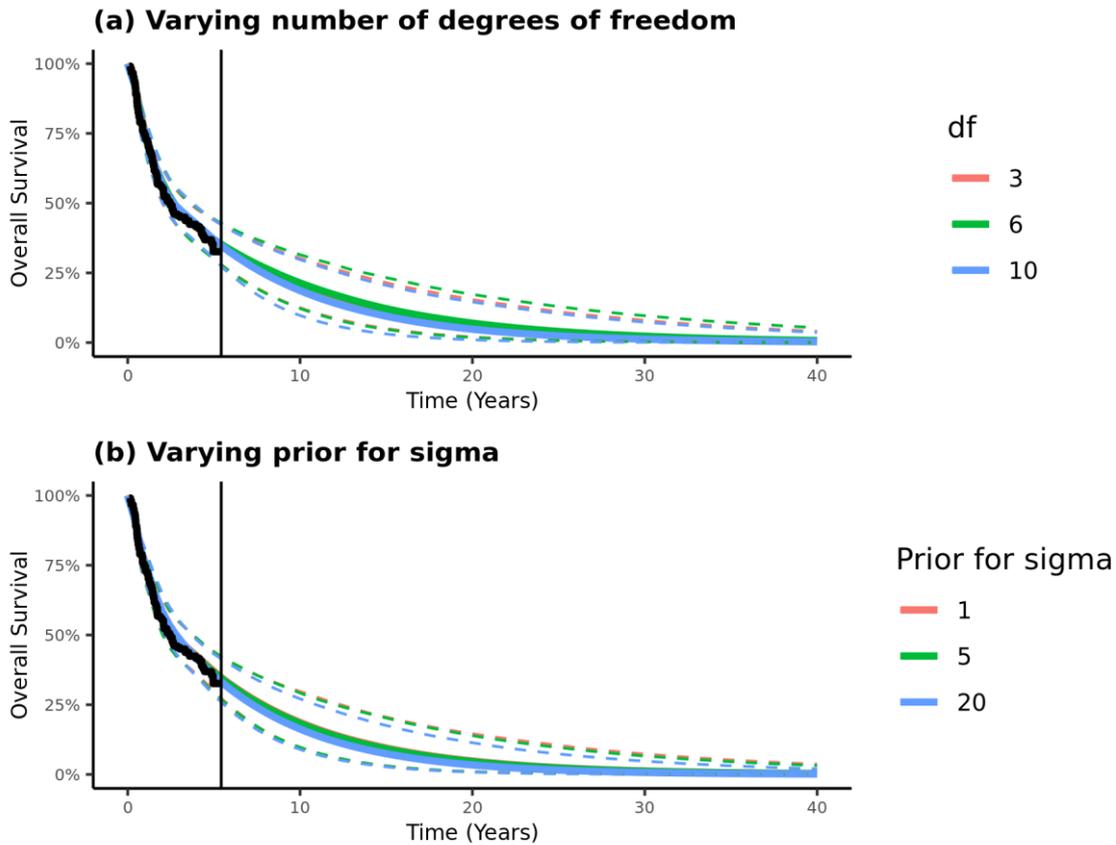

**Supplementary Figure 2:** Survival (posterior median and 95% credible interval) for the radiotherapy control arm from Bonner et al trial with or without inclusion of external data. Sensitivity analyses on position of extra knots in post-trial period.

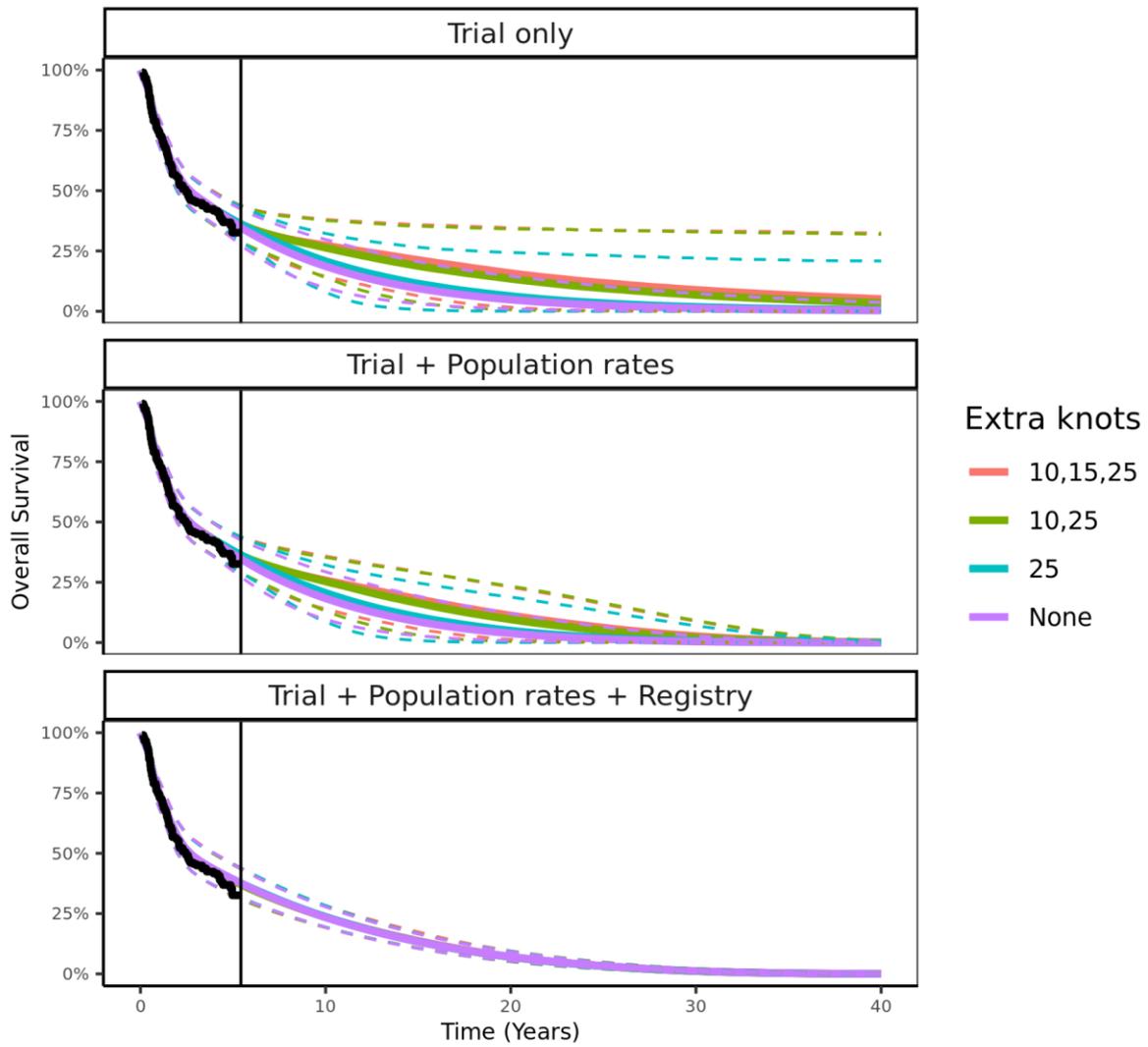

**Supplementary Figure 3:** Posterior hazard samples for models fitted to radiotherapy control arm from Bonner et al trial without inclusion of external data, with different knot configurations. Hazard trajectories are presented for 20 randomly selected posterior iterations from the MCMC sample.

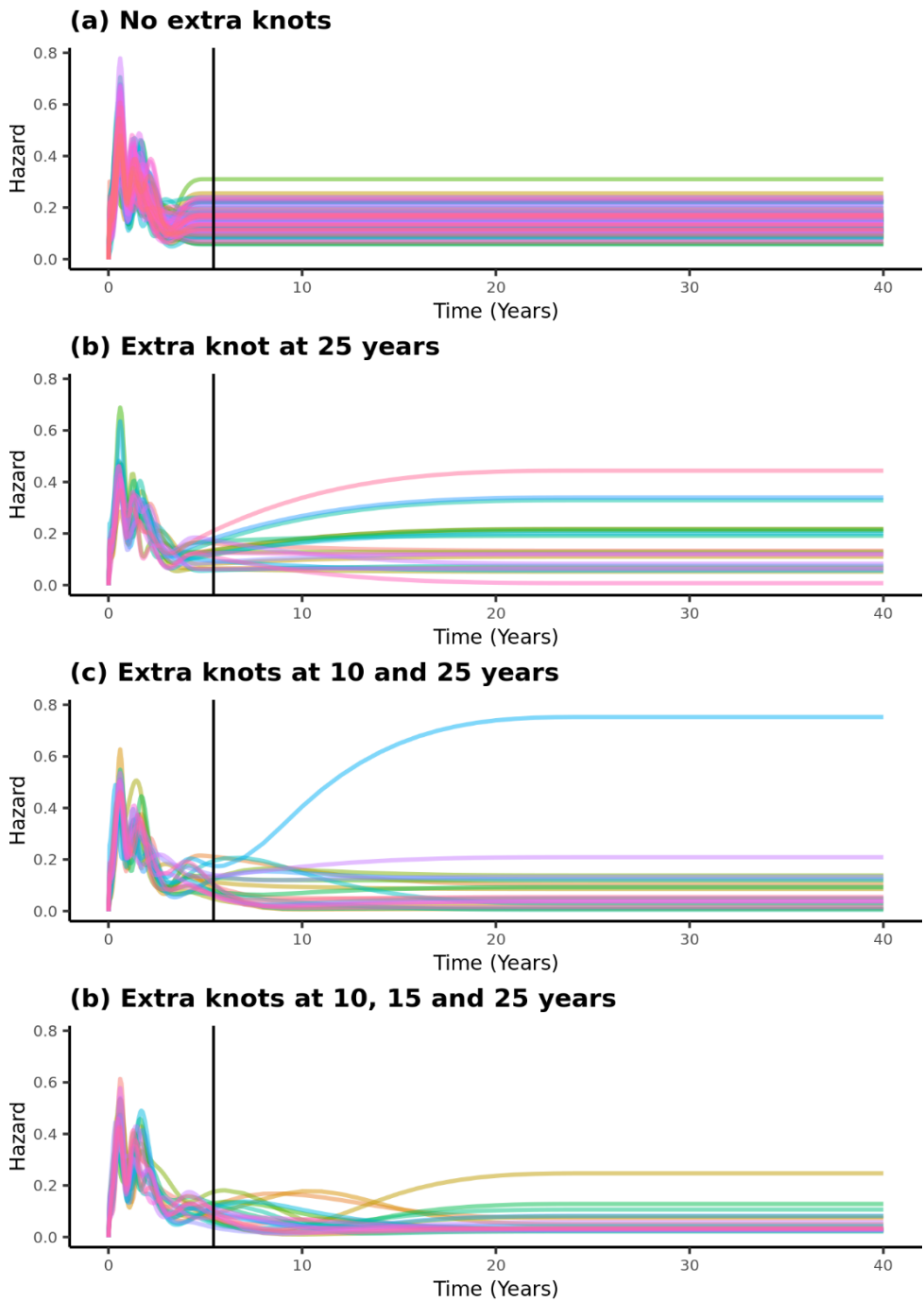

**Supplementary Figure 4:** Hazard ratio (posterior median and 95% credible interval) for the active versus control arm from Bonner et al trial, with sensitivity analysis on modelling choice and extra knot placement.

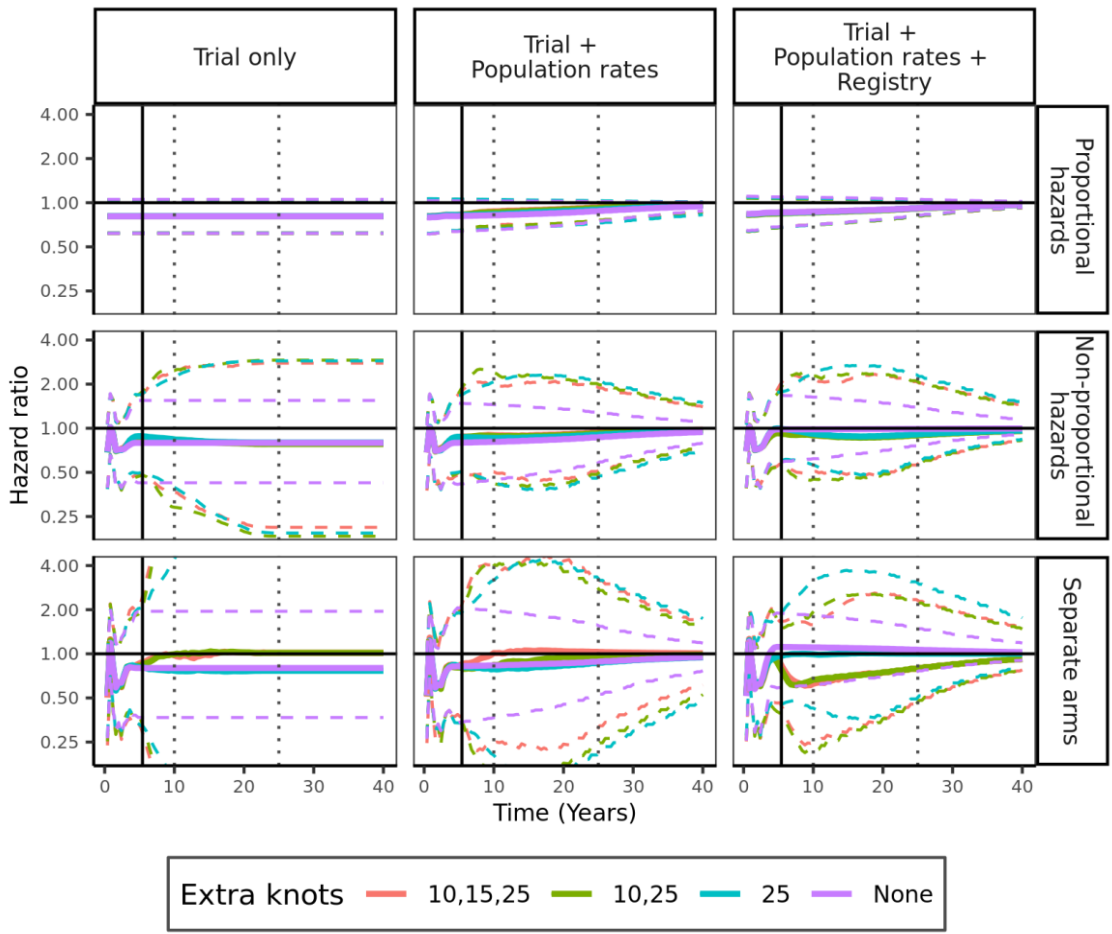

**Supplementary Figure 5:** Bias in marginal 40-year RMST in the control arm when varying trial data cut-off (maturity), placement of extra knots, and inclusion/omission of external data. The vertical line shows the true value.

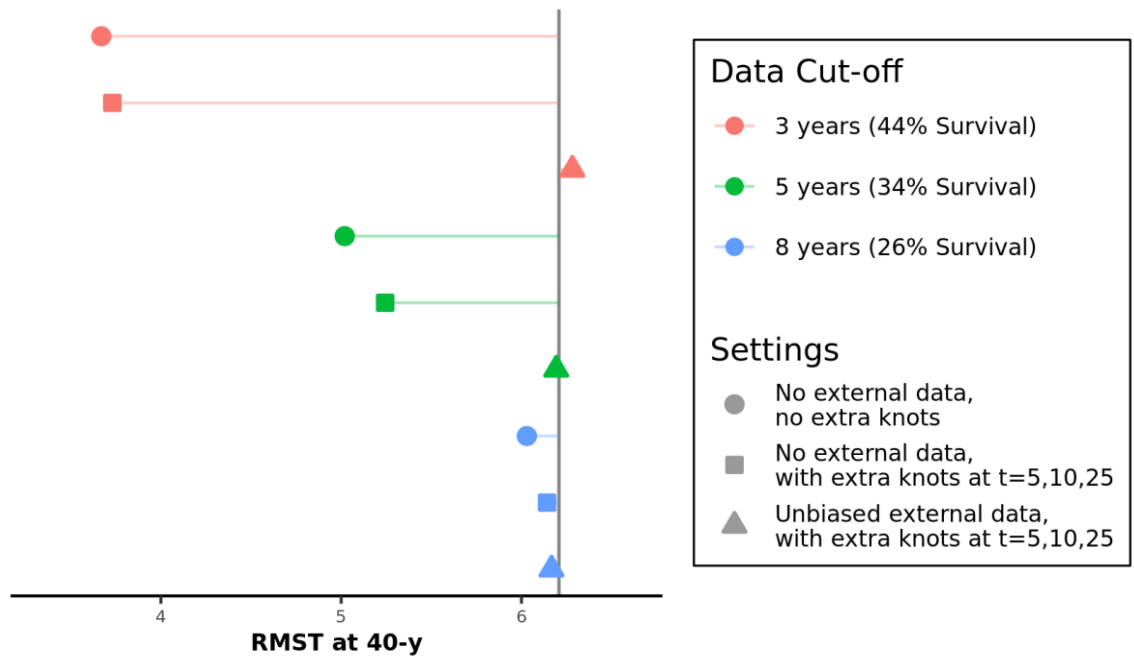

**Supplementary Figure 6:** Bias in (a) marginal 40-year RMST in the control arm, (b), (c), (d) marginal 40-year RMST difference between active and control arms, when varying the number of knots (degrees of freedom (df)) within trial follow-up, the number of extra knots beyond trial follow-up, the inclusion/omission of external data and the treatment effect modelling strategy. In (b), (c) and (d) all models include external data on the control arm. The vertical line shows the true value.

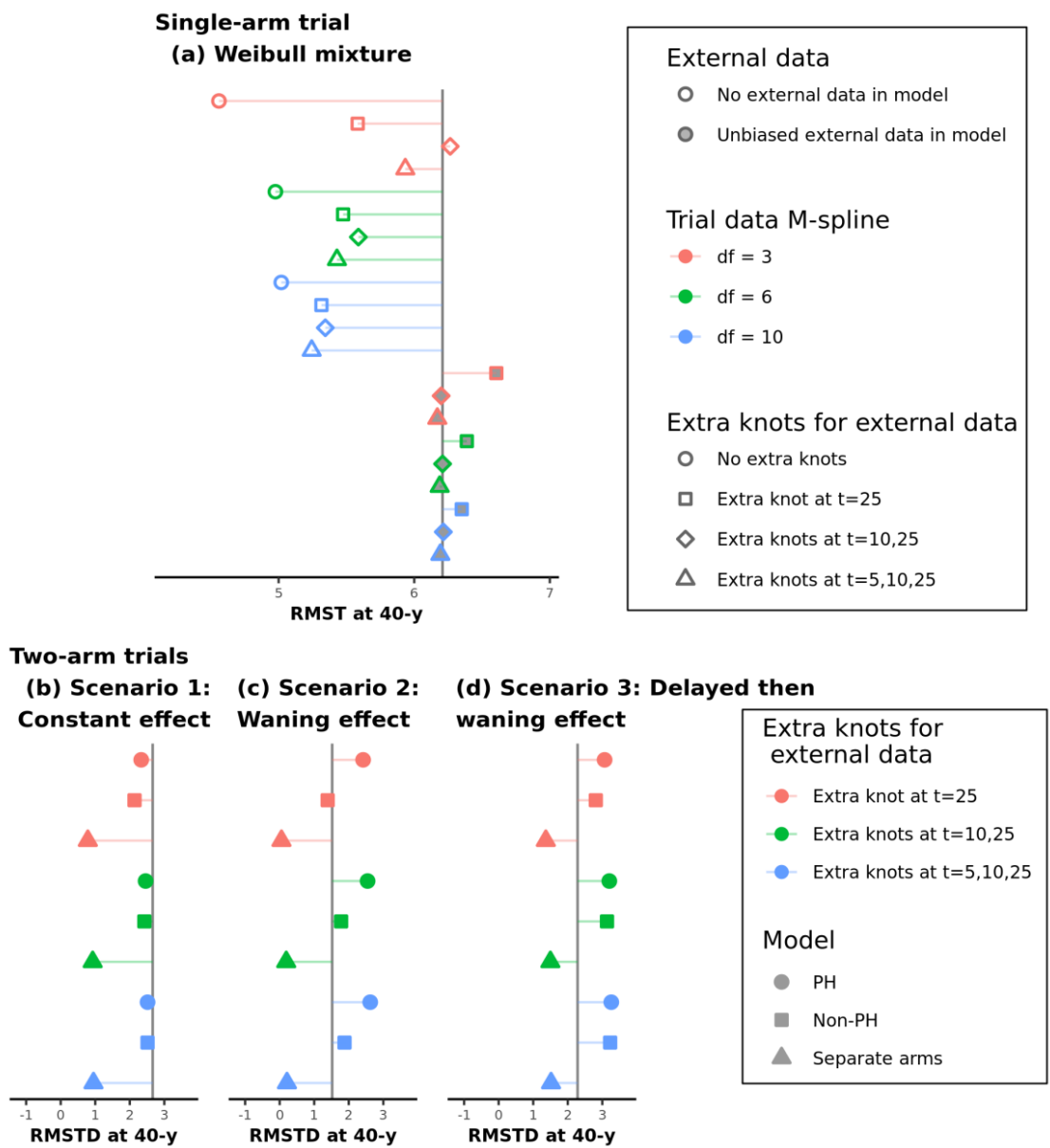

**Supplementary Figure 7:** Two-arm scenarios, individual arm RMST and difference in RMST (RMSTD) at 40-years.

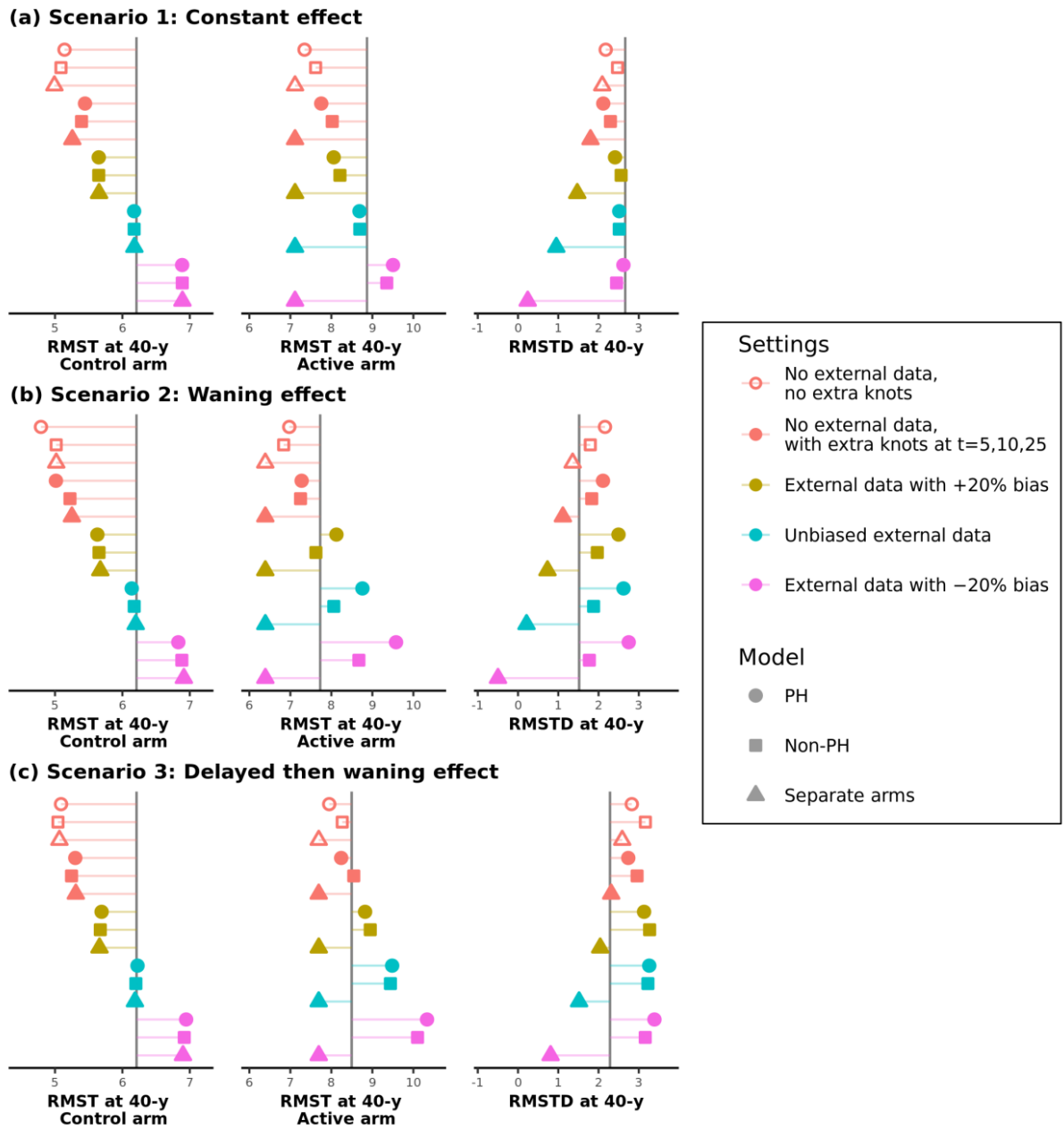

**Supplementary Figure 8:** Bias of RMST differences under proportional hazards and non-proportional hazards models, compared between different true and assumed scenarios of treatment effect waning.

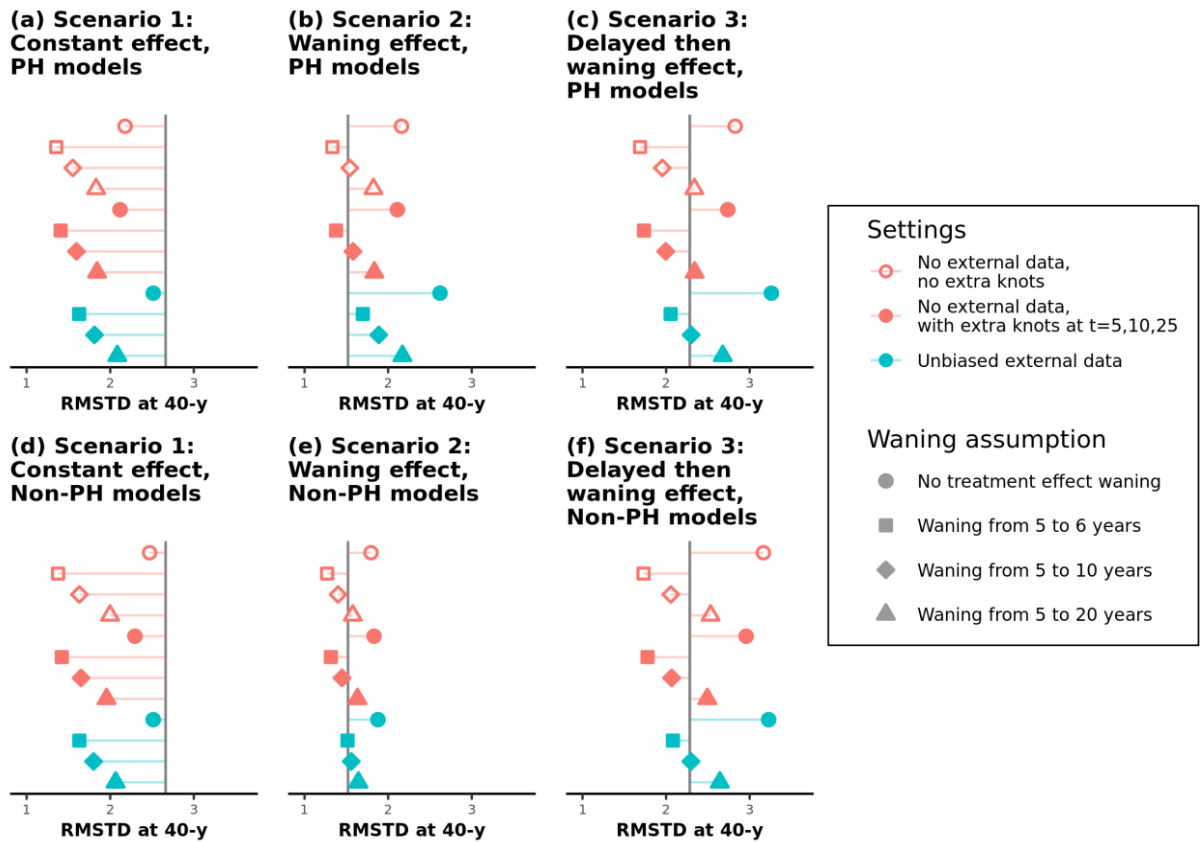

# Supplementary Methods - References